\documentclass{ws-procs10x7}
\def\ie{{\it i.e.}}
\def\eg{{\it e.g.}}

\def\to{\rightarrow}

\begin{document}

%\rightline{\vbox{\halign{&#\hfil\cr
%%&DRAFT\cr
%&SLAC-PUB-10687\cr
%&September 2004\cr}}}
%\begin{center}

\title{Little or No Higgs}

\author{T.G. Rizzo}
%{\footnote {Talk given at the $32^{nd}$ International Conference on High 
%Energy Physics, Bejing, China, 16-22 August 2004}}

\address{Stanford Linear Accelerator Center, 2575 Sand Hill Rd.  
Menlo Park CA 94025,
USA\\E-mail: rizzo@slac.stanford.edu}

\twocolumn[\maketitle\abstract{Both Little Higgs and Higgsless Models 
provide new windows into the mysteries of electroweak symmetry breaking and 
lead to testable predictions at present and future colliders.
Here we give a quick overview of three papers submitted to the ICHEP2004 
meeting on these subjects.}]

\section{Monte Carlo Analysis of the Warped 5D Higgsless Model}%1

Higgsless models offer the possibility of electroweak symmetry breaking without 
the introduction of any explicit additional 
Higgs fields in the action within the extra dimensional 
framework. The various symmetries are broken 
via the application of appropriately chosen boundary conditions(BC) on the 5D 
gauge field wavefunctions. The 
role of the Goldstone bosons is played by the  
additional scalar component of the original 5D gauge fields which are `eaten' 
to generate the massive 4D Kaluza-Klein(KK) tower of gauge fields. Unlike in 
orbifold models the would be zero mode can obtain a finite mass. 

The most phenomenologically 
sucessful framework for a model of this type has been presented 
in Refs.{\cite{csaki,us}}. In these models one takes the usual Randall-Sundrum 
5D warped setup{\cite {rs}} and places a Left-Right Symmetric 
$SU(2)_L\times SU(2)_R\times U(1)_{B-L}$ gauge 
theory in the bulk with the SM fermions being localized to the Planck brane. (The LR 
symmetry helps insure that $\rho$ parameter 
is very nearly unity.) The BC are chosen to break 
$SU(2)_R\times U(1)_{B-L} \to U(1)_Y$ on the Planck brane while simultaneously 
breaking 
$SU(2)_L\times SU(2)_R \to SU(2)_D$, the diagonal subgroup, 
on the TeV brane leaving just $U(1)_Q$. In addition, brane kinetic 
terms for the gauge fields corresponding to the unbroken generators on the 
two branes are also 
present: $SU(2)_D$ and $U(1)_{B-L}$ on the TeV brane and $SU(2)_L$ and $U(1)_Y$ 
on the Planck brane. The relative strength of these brane terms are described via 
a set of dimensionless parameters $\delta_{D,B,L,Y}$, which, together with 
$\kappa=g_R/g_L$, the ratio of the gauge couplings, are the free parameters 
of this model. Other model 
parameters can be traded for $M_{W,Z}$ and $G_F$ in various combinations. 

There are a large 
number of constraints that such models must satisfy and there is a strong 
tension in this model between these various requirements: ($i$) agreement 
with the precision electroweak data (which, since we are performing a tree level 
analysis, we interpret to mean the tree level SM), ($ii$) the bounds on the masses 
and couplings of KK excitations from direct and indirect(\ie, contact interaction) 
searches and ($iii$) perturbative unitarity(PU) in $W_L^+W_L^-$ elastic 
scattering. Recall that without the Higgs the SM violates perturbative unitarity at 
a scale of $\sqrt s \simeq 1.7$ TeV; here the KK's must unitarize 
this scattering amplitude instead of the Higgs via $s-$ and $t-$channel exchanges.  
Naively this amplitude grows like $s^2$ but all such terms are cancelled 
due to gauge invariance when the 
various diagrams are combined  thus leaving an $s$-like 
growth. The SM Higgs cancels these terms leaving an amplitude which has no 
power-like growth in $s$; here in this model the Higgs is absent so that this 
cancellation is non-trivial. In principle, the remaining amplitude that naively 
behaves as $s^0$ can still be large. 

\begin{figure}
\centerline{
\includegraphics[width=5.0cm,angle=-90]{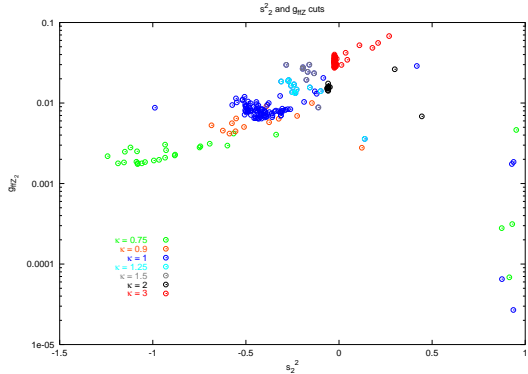}}
\caption{Mass and effective $\sin^2 \theta$ for the first KK excitation beyond 
the $Z$ for models passing the criteria discussed in the text.}
\label{fig1}
\end{figure}

To see if an allowed model 
parameter space exists we{\cite {us}} generated via Monte Carlo techniques 
$\sim 5\cdot 10^6$ sets of models described by the above parameters. We then asked 
whether all the constraints above could be satisfied simultaneously by passing them 
through a series of filters. We first demanded that all $g_i^2>0$ and that there 
be no ghost states in the spectrum. We then 
required the value of $\rho_{eff}$, defined via the $W$ and $Z$ couplings,to be 
within $0.5\%$ of unity and that 
the various electroweak couplings be the same as their SM values at the same level  
of precision, $0.5\%$. 
In our parameterization, deviations from the usual SM 
couplings are described 
by the $\rho_{eff}$ parameter and the existence of three different 
$\sin^2\theta_w$'s: the on-shell value $\sin^2\theta_{OS}=1-M_W^2/M_Z^2$ using the 
observed $W$ and $Z$ masses as input, $\sin^2 \theta_{eg}=e^2/g_W^2$, where $g_W$ 
is defined through the $W$ coupling to the SM 
fermions with $e$ being the electric charge and 
$\sin^2 \theta_{eff}$ defined the the $Z$-pole couplings to fermions. 
We remind the reader 
that this is a tree level analysis and that these three quantities are a priori 
different in this model but are identical in the tree level SM. (They do differ 
in the SM at the loop level.) 
Next all TevII and LEPII constraints on the KK's must be satisfied including 
direct searches for KK resonances and the contact interaction effects induced 
via KK exchanges. Similarly,   
the couplings and masses of the first neutral KK must be such as to 
potentially lead to perturbative unitarity(PU). Certainly if the first neutral 
KK state is more massive than $\simeq 1.5$ TeV 
it will be too heavy to rescue unitarity. In 
addition, if the same KK couples mainly to hyperchage it will not couple to 
$W_L^+W_L^-$ and thus will clearly fail in providing unitarity.
Fig.1 shows that there are a reasonable set of models 
that survive these general requirements. (Note that tightening up the electroweak 
constraints, \eg, letting $0.5\% \to 0.1\%$, reduces the number of survivors by about 
a factor of 10.) Unfortunately, also requiring that 
there be no dangerous tachyons that couple strongly to SM fields removes almost 
all (but 6!) of these survivors. Tachyons of this dangerous type are found to 
exist when $\delta_{B,D}<0$ or when $\delta_L <<0$ which describes most of the 
survivors. 

To examine PU for the remaining models we performed a straightforward evaluation of the 
$J=0$ partial wave amplitude, $a_0$, and required $|a_0|< 1/2$ as usual. Unfortunately 
these models fail this requirement when $\sqrt s \geq 2$ TeV or so which is not 
much better than the SM with a Higgs. It is possible that this straightforward 
application of this now standardized test is too restrictive in the class of models 
under consideration and other tests may be more applicable. This is currently under 
active investigation.

\section{6D Higgsless Model in Flat Space}%2

It is interesting to examine the possible extensions of the 5D Higgsless 
model. 
Perhaps going to 6D may alleviate some of the problems of the 5D warped 
Higgsless model due to greater flexibility; 
the simplest 6D scenario, with only the SM in a flat space 
bulk, was considered in Ref.{\cite {nandi}}. Here the symmetries are broken 
via boundary conditions along the orthogonal additional directions in two 
steps, \ie, $SU(2)_L \to U(1)_{I_3}$, the third component of weak isospin, 
and $U(1)_{I_3} \times U(1)_Y \to U(1)_Q$, 
as is shown in Fig.2. In addition to the two gauge couplings, $g,g'$, and the radii of 
the two additional dimensions, $R_i$, there are two mass parameters in the 6D 
bulk action, $M_{L,Y}$, employed to produce fields and couplings with canonical 4D 
dimensionalities. Besides this bulk action, the conventional 4D SM gauge 
fields plus the SM fermion fields are localized at $(0,0)$ as boundary terms. 
In principle one imagines that $R_1 \sim R_2$ and $M_L \sim M_Y$ so that no 
substantial parameter hierarchy exists. If the brane terms dominate those in the 
bulk then one has the inequality $g^{-2},g'^{-2} >> (\pi M_{L,Y}R_{1,2})^2$ 
which is the model limit examined in Ref.{\cite {nandi}}. 

Note that the masses of the SM $W$ and $Z$ have somewhat different origins, \ie, 
whereas the $W$ mass arises from the BC's only on the $y_1=\pi R_1$ surface, the $Z$ 
mass also obtains contributions from the $y_2=\pi R_2$ surface. To 
leading order, \eg, one obtains $M_W^2=2g^2M_L^2R_2/R_1$ whereas 
$M_Z^2=2(g^2+g'^2)M_L^2M_Y^2(R_1/R_2)/(M_L^2+M_Y^2)$. KK excitation 
masses go as $\sim 1/R_i$, 
\eg, the first $W$ KK excitation has a mass of $1/R_1$ to leading order. These 
mass relationships lead to a fine-tuning of the $\rho$ parameter since to 
leading order 
one obtains $\rho= (1+M_L^2/M_Y^2)R_2^2/R_1^2$ plus higher order terms that have been 
dropped. (To obtain this result the usual 
relation $\cos^2 \theta_w=g^2/(g^2+g'^2)$ has been employed.)  
This ratio is naturally of order unity but 
is $\neq 1$ in general. In order to satisfy the electroweak constraints these set 
of parameters must thus be tuned to $\sim$ 1 part in $10^3$. Once this tuning is made 
it can be shown that the other oblique parameters are suitably small.

\begin{figure}
\centerline{
\includegraphics*[width=6.0cm, bb = 163 588 449 752]{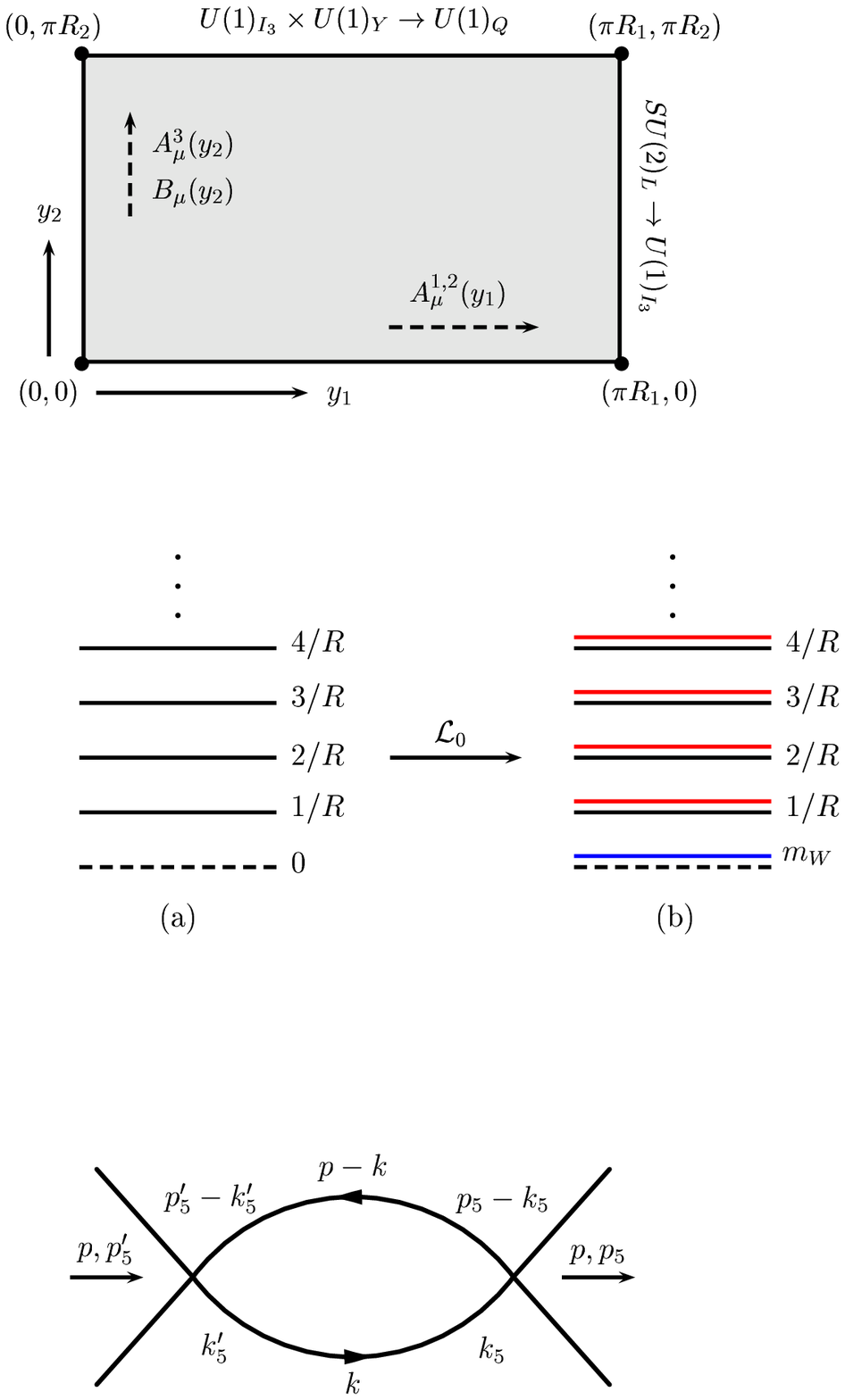}}
\caption{Symmetry breaking by boundary conditions of the $SU(2)_L\times U(1)_Y$ 
SM in 6D flat space.}
\label{fig2}
\end{figure}

In addition to this tuning problem there are a number of other unanswered questions 
about this model: ($i$) The fermions remain massless as they are localized at 
$(0,0)$ 
where the SM remains unbroken. How are fermion masses to be generated? ($ii$) What 
are the couplings of the various KK states to the SM fermions? Are direct and 
indirect collider bounds on the KK spectrum easily satisfied? ($iii$) How are 
issues related to perturbative unitarity in $W_L^+W_L^-$ elastic scattering 
addressed in this model? The answers to these and other questions  
may shed light on the phenomenological validity of this class of models.

\section{Fermion Completion of the Simplest Little Higgs Model}%3

In Little Higgs models the light SM-like Higgs is a psuedo-Goldstone boson which 
is protected from getting a mass by a pair of global symmetries. These same 
symmetries also remove the one-loop quadratic divegences in the Higgs sector that 
are present in the SM though these divergences remain at two-loop order. 
All such models have additional gauge, fermionic and scalar 
degrees of freedom at the TeV scale which play important roles in removing the 
quadratic divergences. In the simplest $SU(3)_L\times U(1)_X$ model of Kaplan and 
Schmaltz{\cite {ks}}, the left-handed top quark becomes a member of an $SU(3)_L$ 
triplet representation, $3_L$, together with $b_L$ and an 
$SU(2)_L$ isosinglet quark with $Q=2/3$, 
$T_L$, of the form $(t,b,T)_L^T$; all RH fields are singlets as in the SM. 
Note that in addition to the $W,Z,\gamma$ of the SM there exists a new neutral 
hermitian gauge boson, as well as a pair of non-hermitian neutral gauge fields connecting 
$t_L$ and $T_L$ and a new pair of charged gauge fields connecting $b_L$ and $T_L$. 
Since the top quark yields the largest fermionic contribution to the Higgs 
quadratic divergence, here cancelled by the contribution of $T$, little attention is 
generally paid as 
to how the other SM fermions transform. This issue has been taken up by 
Kong{\cite {kong}}. 

Kong has shown that there is a unique way to introduce the other SM fermions within 
the $SU(3)\times U(1)$ framework while maintaining anomaly freedom: ($i$) The 
first two generations of LH-quarks are placed in anti-triplets, \ie, $\bar 3_L$'s, 
\eg, $(d,u,D)_L^T$, with their RH-partners remaining as singlets; here $D$ is an 
$SU(2)_L$ isosinglet $Q=-1/3$ state. 
($ii$) All three generations of the LH-leptons are 
placed in triplets of the form $(\nu, \ell,N)_L^T$ while the RH partners of $\ell$'s  
are singlets. Here $N$ is a neutral $SU(2)_L$ isosinglet state that can be 
of use in generating small neutrino masses. The SM-like Higgs fields lie in two 
anti-triplets and lead to interesting fermion mass patterns that can be generated by 
terms originating from dimension-4 and dimension-5 operators. 

Perhaps the most serious issue facing this scenario is the existence of FCNC's 
that arise after fermion mixing 
from a number of sources including the transformation of the new fermions 
into those of the SM via the emission of a $Z$. These occur due to the fact that the 
conventional 
Paschos-Weinberg-Glashow conditions are no longer satisfied. Other FCNC sources also 
appear through 
the exchanges of the $SU(3)_L$ gauge bosons not present in the SM. Whether it is  
possible to satisfy the existing experimental constraints while simultaneously 
satisfying quartic divergence cancellation, precision electroweak data and collider 
bounds remains unknown. This model is in need of further study.

\section{Conclusions}

While it is clear that our understanding of electroweak symmetry breaking is incomplete 
there is a growing number of new ideas coming on the market. Soon, data from the 
LHC and the ILC will help elucidate this situation. Hopefully by the time of ICHEP2008 
we will already have a much clearer view of the physics behind electroweak symmetry 
breaking.

\section*{Acknowledgments}

Work supported in part by the Department of Energy, 
Contract DE-AC02-76SF00515.

\end{document}